\documentclass[12pt]{article}
\usepackage{amsmath,amssymb,latexsym}
\usepackage[mathscr]{eucal}
\usepackage{graphicx}
\usepackage{dcolumn}
\usepackage{bm}

\newcommand{\beq}{\begin{equation}}
\newcommand{\eeq}{\end{equation}}
\newcommand{\beqa}{\begin{eqnarray}}
\newcommand{\eeqa}{\end{eqnarray}}

\newcommand{\uno}{\mbox{1\kern-.59em {\rm l}}}

\newcommand{\be}{\begin{equation}}
\newcommand{\ee}{\end{equation}}
\newcommand{\bea}{\begin{eqnarray}}
\newcommand{\eea}{\end{eqnarray}}

\begin{document}
\begin{titlepage}
\begin{flushright}
{ROM2F/2006/24}\\
\end{flushright}
\begin{center}
\vspace{2.2cm}
{\large \sc Deflected Anomaly Mediation and Neutralino Dark Matter}\\
\vspace{2.2cm}
{\sc Alessandro Cesarini, Francesco Fucito and Andrea Lionetto}\\
\vspace{1cm}
{\sl Dipartimento di Fisica, Universit\`a di Roma ``Tor Vergata''\\
I.N.F.N. Sezione di Roma II,\\
Via della Ricerca Scientifica, 00133 Roma, Italy}\\
\end{center}
\vskip 1.5cm
\begin{center}
{\large \bf Abstract}
\end{center}
{This is a study of the phenomenology of the neutralino dark matter
  in the so called deflected anomaly mediation scenario.
This scheme is obtained from the minimal anomaly mediated scenario by
introducing a gauge mediated sector with $N_f$ messenger fields.
Unlike the former scheme the latter has no tachyons.
We find that the neutralino is still the
LSP in a wide region of the parameter space:
it is essentially a pure bino in the scenario
with $N_f=1$ while it can also be a pure higgsino for $N_f>1$.
This is very different from the naive anomaly mediated scenario
which predicts a wino like neutralino. Moreover we do not find any tachyonic scalars
in this scheme.
After computing the relic density (considering all the possible
coannihilations) we find that there are regions in the parameter
space with values compatible with the latest WMAP results with no need to consider moduli fields that decay in the early universe.}

\par    \vfill
\end{titlepage}
\addtolength{\baselineskip}{0.3\baselineskip}
\section{Introduction}
Dark matter still remains one of the main unsolved problem in
physics. The common accepted paradigm is the existence of an exotic
weakly interacting massive particle (WIMP). Such a particle has to be
found in some extension of the Standard Model (SM) of particle physics. It is well
known that supersymmetry is an essential ingredient of a consistent
theory beyond the SM and the most studied framework
is the MSSM, the minimal supersymmetric extension of the SM. In the
MSSM the lightest supersymmetric particle (LSP) is usually a
neutralino, which is a good candidate for cold dark matter~\cite{Jungman:1995df}.
The pattern of the soft supersymmetry breaking terms\footnote{for a recent review about soft supersymmetry breaking lagrangian see~\cite{Chung:2003fi} and~\cite{Shadmi:2006jv}} greatly affects the composition and the strength of the dominant interactions of the neutralino. Hence it is very interesting to study the neutralino phenomenology in different supersymmetry breaking scenarios.
It is in general very difficult to find a mechanism which is able to generate a suitable soft supersymmetry breaking lagrangian without the need of any fine-tuning.
There are essentially three main classes of possible mechanisms which are differentiated by the vehicle which transmits the supersymmetry breaking from the primary ``source'' to the MSSM fields.
The first and most studied scenario is gravity mediation~\cite{msugra}
in which the supersymmetry breaking is vehicled by tree-level Planck
suppressed couplings. Another widely studied scenario is gauge
mediation (GMSB)~\cite{Giudice:1998bp} in which ordinary gauge
interactions vehicle the breaking. The last scenario is anomaly mediation (AMSB)~\cite{Randall:1998uk} in which supersymmetry breaking is transmitted to the MSSM due to the $R$-symmetry and scale anomalies.
Although very natural, the gravity-mediated schemes have the unwanted feature that in order to be phenomenologically viable they must suppose fine-tuned forms for the superpotential and K\"ahler potential \cite{Randall:1998uk}. On the other hand the gauge-mediated scheme does not have such problem, but it predicts the gravitino as the LSP, which is not the most suitable candidate to constitute dark matter.

The anomaly-mediated models do not have any of the last two undesirable features, since they usually predict the neutralino to be the LSP and any direct gravitational coupling between the primary supersymmetry breaking source and the MSSM to be suppressed.
Anyway the minimal AMSB (mAMSB) scheme predicts some MSSM scalars to be tachyonic and hence there must be some other mechanism to lift their squared masses to positive values.
In this paper we consider AMSB models with an additional GMSB-like
contribution to the soft terms which makes them tachyon free. Such scheme is called deflected anomaly
mediated (dAMSB)~\cite{Okada:2002mv,Rattazzi:1999qg}.
The plan of the paper is as follows: in the second section we give a brief
introduction to the minimal anomaly mediated scenario while in the
third section we describe the deflected anomaly mediated scenario. In
the fourth section we show what kind of supersymmetry breaking terms
arise from deflected anomaly mediation leaving to the appendix all the
details of the computation. In the fifth section we present the phenomenological implications for dark matter. The last section is devoted to the conclusions.

\section{Anomaly Mediation Revisited \label{amrev}}
We consider an expansion of the $D=4$ supergravity action in inverse powers of the Planck mass ($M_P\simeq 1.2\times 10^{19}$ GeV).
We are only interested in terms that have no Planck suppression and that involve only the vierbein and the complex scalar auxiliary field $M$ of the supergravity multiplet, besides the terms which do not contain any supergravity fields at all.
The lagrangian~\cite{Randall:1998uk} can be written as:
\begin{eqnarray}
\frac{L}{e} & = & \left[f(Q^\dagger,e^{-V}Q) \varphi^\dagger
  \varphi\right]_{\theta^2{\bar{\theta}}^2}+\nonumber\\
& & \left[\left(\varphi^3 W(Q)+\tau_{ab}\left(Q\right){{\cal W}^a}{{\cal W}^b}\right)_{\theta^2}+h.c.\right]+\nonumber\\
& &+\frac{1}{6}f(Q^\dagger,e^{-V}Q){\cal{R}}+\left\{\mbox{$b_a$ and $\psi_{m}^{\;\;\alpha}$ terms}\right\} + {{\cal{O}}\left(M_P^{-1}\right)}
\label{lagrrs}
\end{eqnarray}
where the $Q$'s are the chiral matter superfields, $\cal{W}$ is the chiral gauge field strength, $b_a$ and $\psi_{m}^{\;\;\alpha}$ are respectively the vector auxiliary field and the gravitino field while $e$ is the vierbein determinant.
The real function $f$ gives the kinetic term for the $Q$'s superfield while the $\tau$ function gives the normalization for the gauge term. We assume for $\tau$ the minimal form, \emph{i.e.} $\tau_{ab}\propto\delta_{ab}$.

The lagrangian in eq. (\ref{lagrrs}) is written in the flat space superfield notation~\cite{bw}, and the spurion chiral superfield $\varphi$ is taken to be
\begin{equation}
\varphi=1-\frac{M^*}{3}\theta^2=1+F_\varphi \theta^2.
\label{gaugefixing}
\end{equation}
By expanding the function $f$ in powers of $M_P$, and modulo a rigid rescaling of the $Q$ superfields, we can write:
\begin{equation}
f(Q^\dagger,e^{-V}Q)=-3M_P^2+Q^\dagger e^{-V}Q+{{\cal{O}}\left(M_P^{-1}\right)}.
\label{effetausvil}
\end{equation}
Inserting this last equation into eq. (\ref{lagrrs}) and dropping all
the scalar curvature, auxiliary vector and gravitino terms we get
\begin{equation}
\frac{L}{e}=\left[Q^\dagger e^{-V}Q \varphi^\dagger \varphi\right]_{\theta^2{\bar{\theta}}^2}+\left[\left(\varphi^3 W(Q)+\tau\left(Q\right){\cal{W}}{\cal{W}}\right)_{\theta^2}+h.c.\right],
\label{lagrstart}
\end{equation}
This action takes into account the couplings of the matter superfields to the complex scalar auxiliary field $M$ of the minimal supergravity multiplet, which could acquire a nonzero vev in case of supersymmetry breaking.
The lagrangian ${\cal L}$ of eq. (\ref{lagrstart}) can be thought of as the effective lagrangian in the flat space limit substituting $M$ with its supersymmetry breaking vev, $M\rightarrow \left<M\right>$, and the vierbein determinant with its flat space value, $e\rightarrow 1$.
In this way, we have a theory which is manifestly invariant under supersymmetry transformations (being written in superspace notation) and in which supersymmetry turns out to be gauge fixed by the condition on the $\varphi$ superfield derived from~(\ref{gaugefixing}).
This is the mechanism able to transmit the supersymmetry breaking to the $Q$ superfields, which
will be finally identified with the MSSM superfields (on the visible brane).
An important remark is that if the superpotential $Q$ had no explicit
mass scale, \emph{i.e.} $W\sim Q^3$, the $\varphi$ dependence in
eq. (\ref{lagrstart}) could be immediately eliminated through a
superfield rescaling:
\begin{equation}
Q\varphi\rightarrow Q.
\label{qresc}
\end{equation}
In this case there would be no tree level communication of the supersymmetry breaking.
However the situation is different at the one loop level because the $\varphi$ superfield cannot be eliminated through the rescaling~(\ref{qresc}).
To see how the mechanism works let us start from the action
corresponding to eq. (\ref{lagrstart}) with $W=y_0 Q^3$:
\begin{equation}
S=\int d^4 x\left\{\left[Q^\dagger e^{-V}Q \varphi^\dagger \varphi\right]_{\theta^2{\bar{\theta}}^2}+\left[\left(y_0 Q^3 \varphi^3 +\tau\left(Q\right){\cal{W}}{\cal{W}}\right)_{\theta^2}+h.c.\right]\right\},
\label{actstart}
\end{equation}
where $y_0$ is a dimensionless parameter.
The classical scale invariance of the action $S$ can be inferred from the absence of explicit mass parameters.
The action $S$ is also classically invariant under the $R$-symmetry, provided that one assigns suitable $R$-weights to the $Q$ and $\varphi$ superfields.
The action of the $R$-symmetry on a chiral superfield $\Phi=\left(A,\psi,F\right)$ of $R$-weight $w_\Phi$ is defined by
\bea
A&\to& A^\prime =e^{2iw_\Phi \lambda} A\nonumber \\
\psi&\to& \psi^\prime =e^{2i(w_\Phi-1) \lambda} \psi \\
F&\to& F^\prime =e^{2i(w_\Phi-2)\lambda} F\nonumber
\label{rsym}
\eea
for real $\lambda$. The kinetic term in eq.~(\ref{actstart}) is $R$-invariant independently of the $R$-weights of $Q$ and $\Phi$, while the superpotential is invariant if one assigns $w_\varphi=2/3$ and $w_Q=0$.
With this assignment the gauge kinetic term has\footnote{In the notation of \cite{bw}, ${\cal{W}}_\alpha=-\frac{1}{4}{\bar{D}}{\bar{D}}D_\alpha V$ and hence the $R$-weight of ${\cal{W}}{\cal{W}}$ is $+2$, since under $R$-symmetry $\theta\rightarrow e^{2i\lambda}\theta$, ${\bar{\theta}}\rightarrow e^{-2i\lambda}{\bar{\theta}}$.} $w=2$ provided that $w_Q=0$ and hence it is classically $R$-invariant.

In order to cancel the ultraviolet divergences, it is necessary to add counterterms, that introduce at least one explicit mass parameter: the ultraviolet cutoff scale $\Lambda_{UV}$ of the theory. This scale appears also considering the theory as an effective one.

The generic form of the action remains that of eq.~(\ref{actstart}). It has a kinetic term and a superpotential term with the same $\varphi^\dagger \varphi$ and $\varphi^3$ couplings with the auxiliary field of the supergravity multiplet.
In addition to the terms in eq.~(\ref{actstart}) the new kinetic and
superpotential terms contain new regulating pieces in which
the $\Lambda_{UV}$ dependence is explicit. The scale invariance of
the theory is now lost due to the presence of a dimensionful
parameter ($\Lambda_{UV}$), while the $R$-symmetry is preserved
because it is determined by $\varphi$ (the only field with nonzero
$R$-weight) which couples to the kinetic and superpotential terms
in the same way as at tree level\footnote{See appendix B of
\cite{Randall:1998uk} for an explicit example of a situation of
this kind.}. It can be seen~\cite{Randall:1998uk} that after the
rescaling of the superfield of eq.~(\ref{qresc}) the $\varphi$
dependence does not disappear from the lagrangian unlike in the
tree level case. The field $\varphi$ appears together with the
explicit cutoff scale $\Lambda_{UV}$:
\begin{equation}
\Lambda_{UV}\rightarrow \varphi\Lambda_{UV}\quad \mbox{ or }\quad \Lambda_{UV}\rightarrow \varphi^\dagger\Lambda_{UV}.
\label{rulelambdauv}
\end{equation}
Under the rescaling~(\ref{qresc}), both $Q$ and $\varphi$ acquire
an $R$-weight equal to $2/3$ and, at the loop level, the
lagrangian obeys the $R$-symmetry. Introducing the wave function
renormalization for the chiral and gauge superfields the lagrangian
becomes\footnote{Due to the nonrenormalization theorem the
trilinear term does not renormalize: all the renormalization
effects reduce to only wave function renormalizations and no
vertex renormalization (see, \emph{e.g.},
\cite{Martin:1997ns,bw}).}:
\[
L=\left[Z_Q\left(\frac{\mu}{\Lambda_{UV}|\varphi|}\right)Q^\dagger e^{-V}Q\right]_{\theta^2{\bar{\theta}}^2}+
\]
\begin{equation}
+\left\{\left[y_0 Q^3+\tau\left(\frac{\mu}{\Lambda_{UV}\varphi}\right){\cal{W}}{\cal{W}}\right]_{\theta^2}+h.c.\right\}.
\label{lanomalyren}
\end{equation}
If we turn off the coupling to gravity, \emph{i.e.} we fix
$\varphi=1$, the $R$-symmetry is lost. This is a consequence of
the fact that without the coupling to $\varphi$ the $R$-symmetry
is anomalous. In this way we have shown that the $\varphi$ field
cannot be decoupled from the loop level lagrangian through the
rescaling of eq.~(\ref{qresc}) as in the case of the tree level
lagrangian which does not contain explicit mass parameters. This
implies that the supersymmetry breaking effects, which are encoded
in the nonzero vev of the $\theta^2$ component of $\varphi$, are
always transmitted at loop level to the matter and gauge
superfields.

\section{Deflected Anomaly}
In its minimal form the anomaly mediated
scenario~\cite{Randall:1998uk} leads to tachyonic masses for the
sleptons (which do not transform under the $SU(3)$ gauge group).
Many different solutions to this problem has been
proposed~\cite{Giudice:1998xp,Luty:1999cz,Luty:2001jh,Luty:2001zv,Jack:2000cd,Arkani-Hamed:2000xj,Katz:1999uw,Chacko:1999am,Allanach:2000gu,Chacko:2001jt,Chacko:2001km}
but one of the most interesting and elegant relies in considering
an additional gauge mediated sector. By introducing $N_f$
messengers fields the RGE gets modified in such a way to avoid
tachyons at the weak scale.

One of the prediction of the deflected anomaly models is the non
universality of the gaugino masses at the GUT scale (even at the
messenger scale).

Let us start from the anomaly mediated sector. We assume the setup
of~\cite{Randall:1999ee,Randall:1999vf,Luty:2000ec} with two
branes in a $D=5$ space-time where the $5^{\rm th}$ component is
compactified over the orbifold $S^1/Z^2$. The hidden brane is the
source of the breaking of supersymmetry through the
(super)conformal anomaly. Such geometrical setup permits to avoid
any non gravitational coupling between the fields of the AMSB
hidden sector (let us denote them by $\Sigma$) and those of the
MSSM (let us denote them by $Q$). Such couplings could lead to
phenomenologically dangerous flavor and CP violating effects.
Their absence is due to a suppression factor, given by the bulk
separation, which multiplies any non gravitational interaction
term between an hidden sector superfield and a visible one. Such
factors are absent for couplings which arise in a four dimensional
space-time.

For example, a K\"ahler potential defined on the hidden brane
\beq
\frac{1}{M^2}\Sigma^\dagger \Sigma Q^\dagger Q
\eeq
is gravitationally rescaled (actually this is a gravitational
redshift) as
\begin{equation}
\frac{1}{M^2}\Sigma^\dagger \Sigma Q^\dagger Q e^{-m_B/\mu_c},
\label{dangerouscoupling}
\end{equation}
where $m_B$ is the mass of the non gravitational bulk state that vehicles the interaction across the bulk.
The exponential suppression factor eliminates all the
phenomenologically dangerous couplings of the form of
eq. (\ref{dangerouscoupling}), once one assumes that every
\emph{non gravitational} bulk state has a mass quite larger than the
compactification scale $\mu_c$.
We do not need to know the detailed dynamic of the AMSB hidden sector
except that the complex auxiliary field $M$ of the four-dimensional
supergravity multiplet must acquire a vev $\left<M\right>$ in order to break
supersymmetry on the visible brane.
Now let us take into account the presence of an extra gauge mediated
sector on the visible brane.
We assume that in this sector there is a gauge singlet chiral
superfield $X=\left(A_X,\Psi_X,F_X\right)$ which is directly coupled
to $N_f$ copies of messenger chiral superfields $\Phi_i$ and
${\bar{\Phi}}_i$, transforming under the fundamentals
and anti-fundamentals of the standard MSSM gauge groups.
The tree level lagrangian for the $X$ superfield, in a $M_P^{-1}$ expansion, is of the same kind of the one in eq. (\ref{lagrstart}), without the gauge kinetic part and with the substitution $Q\rightarrow X$.
We can write:
\begin{equation}
L_X=\left[X^\dagger X\varphi^\dagger\varphi\right]_{\theta^2{\bar{\theta}}^2}+\left\{\left[\left(\lambda_{ij}\bar{\Phi}_iX\Phi_j+W\left(X\right)\right)\varphi^3\right]_{\theta^2}+h.c.\right\},
\label{lx}
\end{equation}
where we explicitly separate the part of the superpotential that
involves only the messenger superfields, from the part that depends
only on $X$.
The coupling of the $X$ superfield to
$\varphi$ ensures that the GSMB hidden sector
is gravitationally coupled to the supersymmetry breaking source $\left<M\right>$.
At this stage supersymmetry breaking can be transmitted to the GMSB
hidden sector at tree level or at one loop level.
The former case corresponds, as it was outlined in section~\ref{amrev},
to a superpotential $W\left(X\right)$ that contains explicit mass
scales~\cite{Okada:2002mv}, while the latter implies $W\left(X\right)\sim X^3$ or zero~\cite{Rattazzi:1999qg}.

In this paper we consider a very general scenario in which the
superpotential of the gauge hidden sector $W\left(X\right)$ contains all the terms with couplings of positive or vanishing mass dimension:
\begin{equation}
W\left(X\right)=c_1 X^3+c_2 \left<F_\varphi\right>X^2+c_3 \left<F_\varphi\right>^2 X+c_4 \left<F_\varphi\right>^3,
\label{wx}
\end{equation}
where the $c_i$ ($i=1,\cdots,4$) are real numbers and $\left<F_\varphi\right>=-\left<M^*\right>/3$.
We also assume, without loss of generality, $\left<F_\varphi\right>$
to be real. In fact this condition could always
be satisfied modulo a rigid rotation of the supergravity auxiliary
field $M$.

As in the usual gauge mediation the messenger superfields $\Phi_i$ and
$\bar{\Phi}_i$ acquire masses of order $\left<A_X\right>$ and a mass
splitting of order $\sqrt{\left<F_X\right>}$ where $\left<A_X\right>$ and $\left<F_X\right>$
are respectively the vevs of the scalar and auxiliary
part of the $X$ superfield. It is exactly the presence of an
intermediate threshold given by the $X$ superfield vev which changes
the renormalization group equations of the soft terms off the AMSB
trajectory. In this way the negative squared masses are no longer present in the spectrum.
The main source of supersymmetry breaking is the $\varphi$
superfield\footnote{Indeed we never have to rely on the exact
  mechanism which generates $\left<M\right>\ne 0$.}
We choose the form of the superpotential in such a way to get the
right vev for the $X$ superfield. In this way, besides a source of
susy breaking for the AMSB we also have a similar one for the GMSB.

Upon substituting $\varphi=1+F_\varphi\theta^2$
into~(\ref{lx}) and dropping all the terms involving the messenger
superfields we have
\begin{eqnarray}
L_X & = &\left(F_\varphi^2A_X A_X^*+F_X F_X^* -\partial_\mu A_X^*
  \partial^\mu A_X-i{\bar{\psi}}_X{\bar{\sigma}}^\mu \partial_\mu
  \psi_X\right)+\nonumber\\
& &+A_X F_X^* F_\varphi+3W(A_X)F_\varphi+F_X\frac{\partial
  W(A_X)}{\partial A_X}+\nonumber\\
& &-\frac{1}{2}\psi_X\psi_X\frac{\partial^2 W(A_X)}{\partial^2 A_X}+h.c.,
\label{lxcompo}
\end{eqnarray}
where from now on, for the sake of simplicity, $F_\varphi$ stands
for $\left<F_\varphi\right>$. The lagrangian~(\ref{lxcompo})
contains a canonically normalized kinetic terms for the complex
scalar $A_X$ and for the Weyl fermion $\psi_X$. The equation of
motion of the auxiliary field $F_X$ is
\begin{equation}
F_X=-A_X F_\varphi -\frac{\partial W^*(A_X^*)}{\partial A_X^*},
\label{eqfx}
\end{equation}
Substituting  eq. (\ref{eqfx}) in eq. (\ref{lxcompo}) leads to the
on-shell lagrangian:
\begin{eqnarray}
L_X & = &-\partial_\mu A_X^* \partial^\mu A_X-i{\bar{\psi}}_X{\bar{\sigma}}^\mu \partial_\mu \psi_X+\nonumber\\
& & -\frac{1}{2}\psi_X\psi_X\frac{\partial^2 W(A_X)}{\partial^2
  A_X}-\frac{1}{2}{\bar{\psi}}_X{\bar{\psi}}_X\frac{\partial^2
  W^*(A_X^*)}{\partial^2 A_X^*}+\nonumber\\
& & -V(A_X,A_X^*).
\label{lxonshell}
\end{eqnarray}
where $V(A_X,A_X^*)$ is the scalar potential which in general depends
from the superpotential coefficients $c_i$.
In our analysis we do not need to know
the precise form of this scalar potential.
We only assume that the scalar field $A_X$ acquires a real vev induced by
$F_\varphi$
\beq
\left<A_X\right>=\xi F_\varphi=m
\eeq
where $\xi$ is an adimensional parameter and $m$ denotes the typical messenger scale.
It is possible to compute the induced vev for $F_X$ with
eq.~(\ref{eqfx}) substituting the corresponding vevs for $A_X$ and
$F_\varphi$:
\begin{equation}
\left<F_X\right>=-(\xi+2c_2\xi+3c_1\xi^2+c_3)F_\varphi^2,
\label{fxvevmixed}
\end{equation}
We also need to compute the mass of the
field $\Psi_X$ in order to ensure that the particle associated to this
field was not the LSP.
From the lagrangian~(\ref{lxonshell}) we immediately read off the mass term
\begin{equation}
m_{\Psi_X}=\langle \frac{\partial^2 W(A_X)}{\partial^2 A_X}\rangle=\left(6c_1\xi+2c_2\right)F_\varphi.
\label{mpx}
\end{equation}


\section{Soft Supersymmetry Breaking Terms \label{ssbt}}
In this section we describe the pattern of the supersymmetry breaking
terms that arises in the deflected anomaly scenario.
The MSSM supersymmetry breaking lagrangian can be written as
\begin{equation}
L_{soft}=-m^2_{ij}{\tilde{q}}_i^* {\tilde{q}}_j-\left[\frac{1}{2}M_\lambda \lambda_a\lambda_a+\frac{1}{2}b_{ij}{\tilde{q}}_i{\tilde{q}}_j+\frac{1}{6}a_{ijk}{\tilde{q}}_i{\tilde{q}}_j{\tilde{q}}_k+h.c.\right],
\label{lsoft}
\end{equation}
where the $\lambda_a$'s denote the gaugino fields for the MSSM
gauge groups transforming in the adjoint representation of the
gauge group, and the ${\tilde{q}}$'s stand for the scalar
components of the various MSSM chiral superfields. $i,j,k$ are
family indices. We have considered the case of an $R$-symmetric
lagrangian (eq. (\ref{actstart})), which does not have any
explicit mass scale in the tree level superpotential and that is
coupled to supergravity through the chiral superfield $\varphi$.
We have shown that upon rescaling the superfield of eq.
(\ref{qresc}) the $\varphi$ dependence can be dropped at tree
level. On the other side, if one deals with the one loop level
action, the $\varphi$ superfield cannot be rescaled away and the
quantum lagrangian, after eq. (\ref{qresc}) is applied, has the
form of eq. (\ref{lanomalyren}). The lagrangian
(\ref{lanomalyren}) describes a theory with no intermediate mass
scales between the renormalization scale $\mu$ (which is to be
thought of the order of the highest MSSM mass) and the UV cutoff
scale $\Lambda_{UV}$. The case without intermediate energy scales
corresponds to the ``standard'' anomaly mediation of
\cite{Randall:1998uk}. In the deflected anomaly scenario the
presence of the GMSB-like hidden sector gives an intermediate
energy scale $\left<A_X\right>=\xi F_\varphi$, where $\xi$ is a new
dimensionless parameter that sets the typical mass of the
messenger superfields. In the presence of an intermediate
threshold the lagrangian~(\ref{lanomalyren}) becomes
\begin{eqnarray}
L&=&\left[Z_Q\left(\frac{\mu^2}{X X^\dagger},\frac{XX^\dagger}{\Lambda_{UV}^2\varphi\varphi^\dagger}\right)Q^\dagger e^{-V}Q\right]_{\theta^2{\bar{\theta}}^2}+\nonumber\\
& & +\left\{\left[y_0 Q^3+\tau\left(\frac{\mu}{X},\frac{X}{\Lambda_{UV}\varphi}\right){\cal{W}}{\cal{W}}\right]_{\theta^2}+h.c.\right\},
\label{linizmixed}
\end{eqnarray}
It is worth noting that the $\varphi$ dependence enters only in terms
with an explicit mass scale, for example $\Lambda_{UV}$.

The usual soft supersymmetry breaking terms depend, besides from
the terms in~(\ref{gaugefixing}), from
\begin{equation}
\left<X\right>=m(1+\theta^2 f/m)
\end{equation}
where we have defined
\begin{eqnarray}
m & = & \left<A_X\right>=\xi F_\varphi\\
f & = & \left<F_X\right>=d\xi F_\varphi^2
\label{mf}
\end{eqnarray}
The parameter $d$ indicates how much the RG anomaly mediated trajectory is deflected
\begin{equation}
\frac{f}{m}=d F_\varphi
\end{equation}
The deflection parameter depends on the superpotential parameters
\beq
d=-\frac{\xi+2c_2\xi+3c_1\xi^2+c_3}{\xi}
\eeq
 It is possible to consider both the scenarios with $d<0$ and $d>0$. The
scenario in which $d<0$ has been explored
in~\cite{Rattazzi:1999qg} and it predicts the LSP to be the
fermionic component $\Psi_X$. Thus in our phenomenological
analysis we assume from now on $d>0$. This scenario is usually
termed as positively deflected anomaly
mediated~\cite{Okada:2002mv}. In order to obtain the expression
for the soft terms one has to put vevs into eq.~(\ref{linizmixed})
and then expand in $\theta,\bar{\theta}$ powers the wave function
renormalizations $Z_Q$ and $\tau$. After the rescaling of the
superfields $Q$ and ${\cal{W}}$, in order to have their kinetic
terms canonically normalized, we are able to read off the soft
supersymmetry breaking terms starting from the lagrangian
\begin{eqnarray}
L& = &\left.Z_Q\left(\frac{\mu^2}{m^2(1+\theta^2 \frac{f}{m})(1+{\bar{\theta}}^2\frac{f}{m})},\frac{m^2(1+\theta^2 \frac{f}{m})(1+{\bar{\theta}}^2\frac{f}{m})}{\Lambda_{UV}^2(1+\theta^2 F_\varphi)(1+{\bar{\theta}}^2 F_\varphi)}\right)Q^\dagger Q\right|_{\theta^2{\bar{\theta}}^2}+\nonumber\\
& &+\left\{\left[y_0 Q^3+g^{-2}\left(\frac{\mu}{m(1+\theta^2
        \frac{f}{m})},\frac{m(1+\theta^2
        \frac{f}{m})}{\Lambda_{UV}(1+\theta^2
        F_\varphi)}\right){\cal{W}}{\cal{W}}\right]_{\theta^2}+\right.\nonumber\\
& &\left. +h.c.\right\},
\label{lll}
\end{eqnarray}
and by matching the result with the lagrangian~(\ref{lsoft}).
In~(\ref{lll}) we have expanded the $e^{-V}$ factor appearing in
the chiral kinetic term and kept only the zero-th order term. We
have also taken into account that the real part of the gauge wave
function renormalization is proportional to $g^{-2}$, where $g$ is
the running gauge coupling. The details of the computation of the
soft supersymmetry breaking terms are contained in
appendix~\ref{eotst} and~\ref{eeftst}.

It is interesting to recover the AMSB case. This limit corresponds to $d=1$:
\begin{equation}
\frac{f}{m}=\frac{\left<F_X\right>}{\left<A_X\right>}=F_\varphi,
\label{decoupling}
\end{equation}
and the lagrangian becomes
\begin{eqnarray}
L&=&\left.Z_Q\left(\frac{\mu^2}{m^2(1+\theta^2 F_\varphi)(1+{\bar{\theta}}^2 F_\varphi)},\frac{m^2}{\Lambda_{UV}^2}\right)Q^\dagger Q\right|_{\theta^2{\bar{\theta}}^2}+\nonumber\\
& &+\left\{\left[y_0 Q^3+g^{-2}\left(\frac{\mu}{m(1+\theta^2
        F_\varphi)},\frac{m}{\Lambda_{UV}}\right){\cal{W}}{\cal{W}}\right]_{\theta^2}+\right.\nonumber\\
& & \left.+h.c.\right\}.
\label{llldec}
\end{eqnarray}
Thus in this limit every effect depending on the high energy theory above the
messenger scale $m$ completely decouples and the low energy theory is
completely UV insensitive.

\section{Lightest Neutralino and Relic Density \label{pheno}}
In this section we examine the low energy predictions of this scenario.
The boundary conditions for the soft terms are given at the
renormalization scale $\mu=m=\xi F_\varphi$ because of the simple form
assumed by the RGEs at this scale (see eqs. (\ref{mgauginiexpl}),
(\ref{trilinearexpl}) and (\ref{mscalexpl}) in the appendix~\ref{eeftst}).
We start with the soft breaking parameters at $\mu=m$ and run them down to the weak scale $M_Z$, by using the appropriate renormalization group equations at two loop level \cite{Martin:1993zk}.
To perform the running we used the ISASUGRA RGE code, which is contained in the ISAJET package \cite{Paige:2003mg}.
For the computations of all the quantities at the weak scale we used
the DarkSUSY code \cite{Gondolo:2004sc}.

As we have already seen the soft term expressions are entirely determined by the two mass parameters $F_\varphi$ and $f/m$ and by the dimensionless number $N_f$.
It is then possible to study the phenomenological properties of this scenario through contour plots in the
$\left(f/m,F_\varphi\right)$ plane.
The scale at which the boundary conditions are given ($m=\xi
F_\varphi$) is determined by fixing $\xi$. The ratio between of the
two Higgs vevs $\tan \beta$, the sign of the Higgs $\mu$ term and the
number of messenger flavors $N_f$ are fixed as well.
Since $F_\varphi$ and $\left(f/m\right)$ are our only independent
parameters we can explore scenarios with different values of the
deflection parameter $d$.

The main result is that the neutralino is the LSP in a wide
portion of the parameter space. This is a consequence of the fact
that the fermionic component $\Psi_X$ of the hidden sector scalar
superfield $X$ has a mass of order $F_\varphi$ (see
eq.~\ref{mpx}). It is in fact always possible to choose the
superpotential coefficients $c_i$ in such a way to make the
$\Psi_X$ arbitrarily heavy. Moreover the gravitino mass is
$m_{3/2}= F_\varphi$ and all the soft breaking masses are
suppressed by the square of the gauge
couplings~\cite{Randall:1998uk}. Thus neither the fermionic
component $\psi_X$ of the gauge singlet superfield (as
in~\cite{Rattazzi:1999qg}) nor the gravitino are the lightest
supersymmetric particle. The neutralino turns out to be bino-like
in almost all of the parameter space when only one messenger field
is present, while for $N_f\ge 2$ there are regions in the
parameter space in which the neutralino is a very pure higgsino.

\begin{figure}[t]
\centering
\includegraphics[scale=0.95]{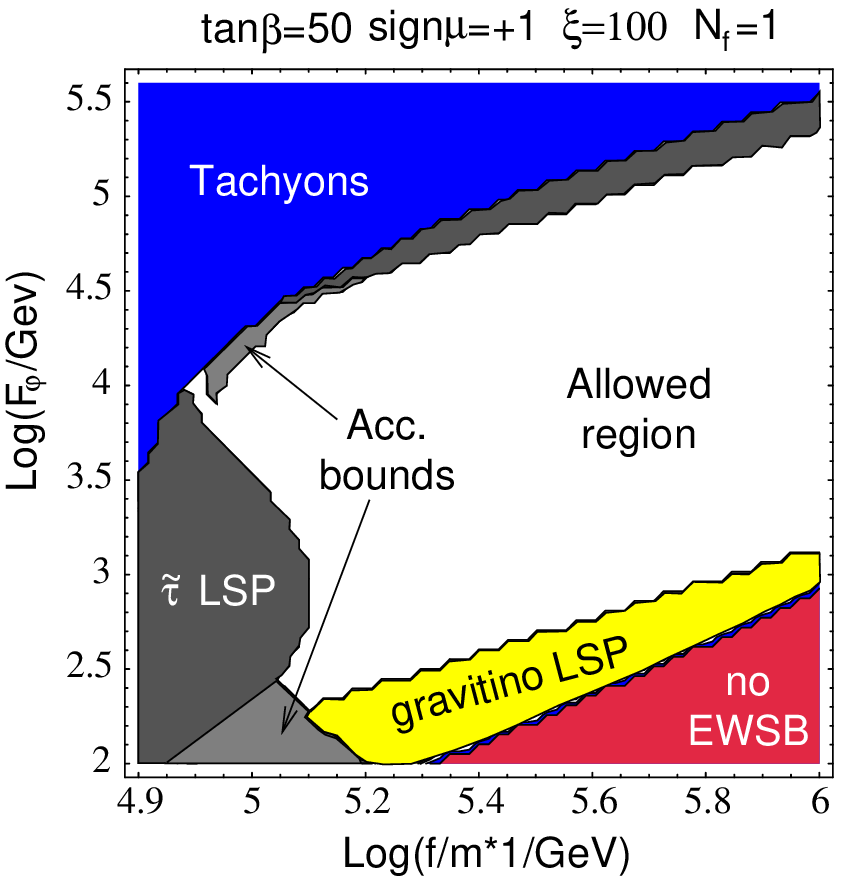}
\includegraphics[scale=0.95]{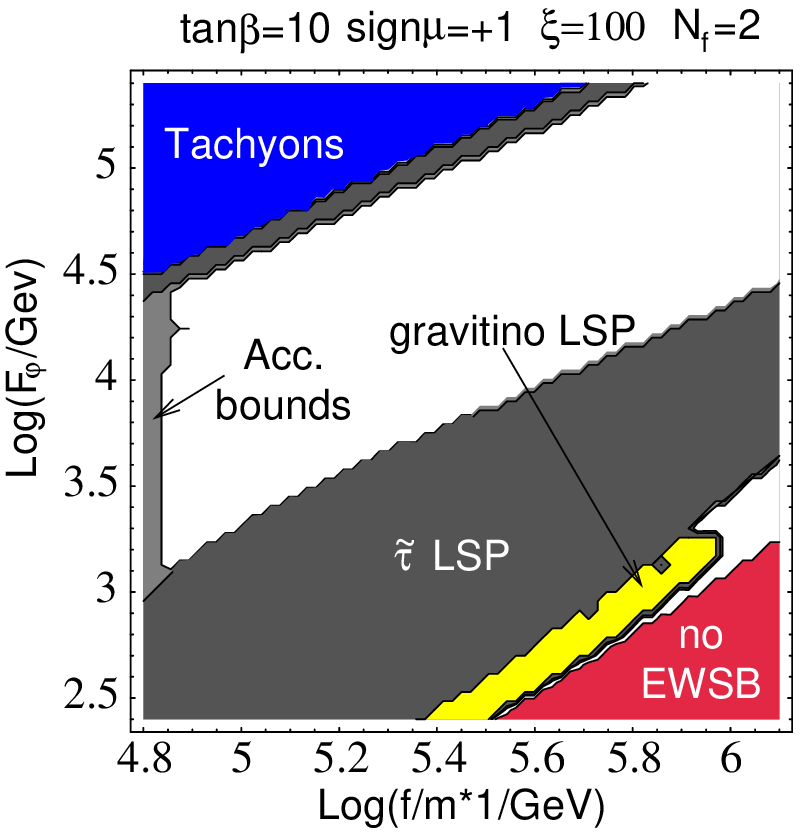}
\caption{Excluded and allowed regions in the plane $(f/m,F_\varphi)$}
\label{fig:excl_nf1_nf2}
\end{figure}

In fig.~\ref{fig:excl_nf1_nf2} we show the regions in the parameter
space already excluded on
phenomenological ground, the regions in which the neutralino is not
the LSP (either gravitino or stau LSP) and the allowed
regions for two different scenarios. In the left panel we fixed $\tan\beta=50$,
$N_f=1$ and the messenger scale to $\xi=100$. The
upper left regions (blue shaded region) is excluded due to the
presence of tachyons in the spectrum (that is the minimal AMSB
result) while the lower right region (red shaded region) is excluded due to an incorrect
electroweak symmetry breaking (EWSB). The yellow shaded region is the
region in which the gravitino is the LSP and it is determined by the
condition
\beq
F_\varphi\leq m_\chi.
\eeq
In the dark gray shaded region the lightest stau
is the LSP. The light dark shaded region is excluded by the current
accelerator constraints on the Higgs boson masses,
$b\to s \gamma$, slepton masses, etc. In particular we considered the
LEP2 lower bound~\cite{Yao:2006px} for the mass of the lightest SM-like Higgs boson $h_0$
\beq
m_{h_0}\geq 114.4\;\; {\rm GeV}
\eeq
In the right panel of fig.~\ref{fig:excl_nf1_nf2} we show the excluded
region for a scenario with $N_f=2$ messenger fields. In this case the
tau LSP region is much wider and a new branch of an allowed region
opens toward higher values of $f/m$. We will see in the following
discussion that this branch is interesting from the point
of view of the neutralino. In general, scenarios with $N_f\ge
1$ are less constrained by the current accelerator data.

There are only slight changes in the excluded and allowed regions for
models with $\xi\lesssim 10$. In general the regions that contains
tachyons and with no EWSB are much larger than the
$\xi\simeq 100$ case.
Models with $\xi\gtrsim 1000$ exhibit a much wider
region in which the gravitino is the LSP and a smaller region with no EWSB.

\begin{figure}[t]
\centering
\includegraphics[scale=0.95]{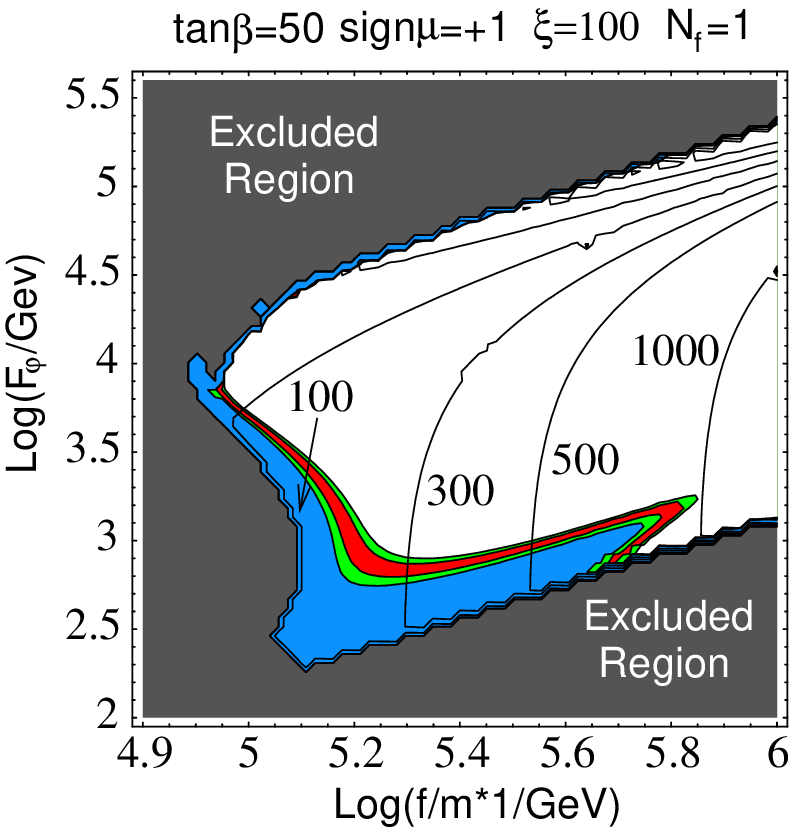}
\includegraphics[scale=0.95]{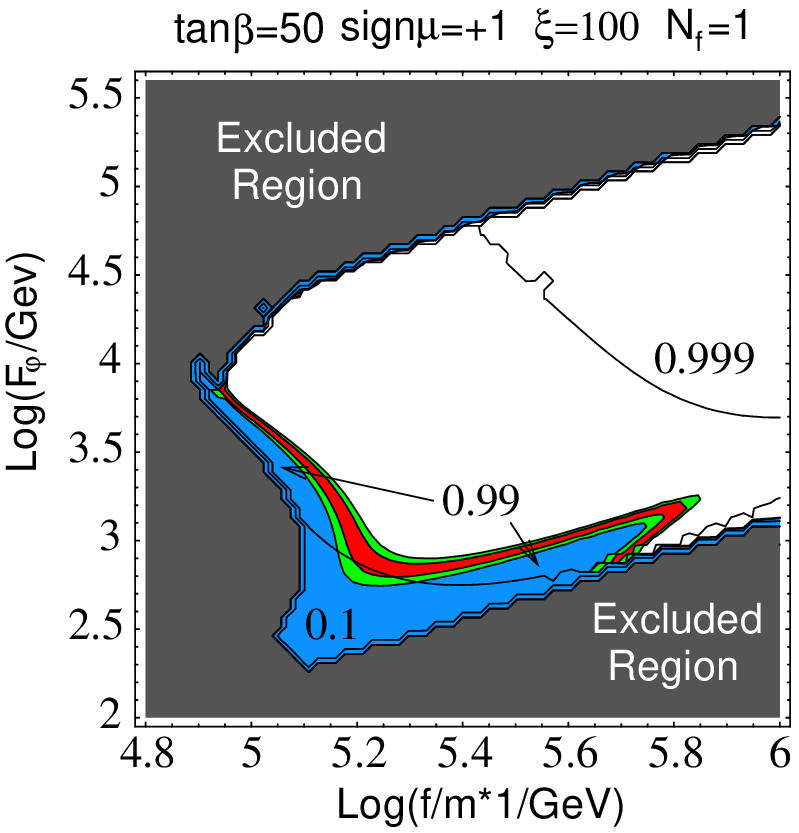}
\caption{Contour plots in the plane
  $(f/m,F_\varphi)$. Left panel: neutralino mass (expressed in GeV). Right panel: gaugino fraction.}
\label{fig:isomchi_zg_tb50_nf1}
\end{figure}

We also computed the thermal relic density $\Omega_{\chi} h^2$ for the neutralino solving the
Boltzmann equations in a standard cosmological scenario. We leave the
analysis of some non standard cosmological scenario involving the
presence of the moduli field associated to the supersymmetry breaking parameter
$F_\varphi$ for future work.
The lightest neutralino is given by the linear combination
\beq
\tilde{\chi}_1^0=N_{11}\tilde{B}+N_{12}\tilde{W}+N_{13}\tilde{H}_u+N_{14}\tilde{H}_d
\eeq
where $\tilde{B}$ and $\tilde{W}$ are the bino and wino fields while $\tilde{H}_u$ and $\tilde{H}_d$ are the two higgsinos. We also define the gaugino fraction as
\beq
Z_g=\left|N_{11}\right|^2+\left|N_{12}\right|^2
\eeq
We say that a neutralino is gaugino-like (in particular in our case bino-like) if $Z_g>0.9$ while is higgsino-like when $Z_g<0.1$. In all the intermediate cases we denote the neutralino as mixed-like.

We show the results for the $N_f=1$ scenario in
fig.~\ref{fig:isomchi_zg_tb50_nf1}. In the left panel we show the
neutralino isomass contours together with the cosmologically favorite
regions: models in the red shaded region have a relic density in the
$2\sigma$ WMAP~\cite{Spergel:2006hy} range $\Omega_{CDM}h^2=0.110\pm
0.014$ while models in the green region are in the $5\sigma$
range with $\Omega_{CDM}h^2=0.110\pm 0.035$.
The blue region denotes models in which the neutralino is really a
subdominant dark matter component with $\Omega_{\chi}h^2\lesssim
0.07$. The allowed neutralino masses range from $m_\chi\simeq 100$ GeV
up to $m_\chi\simeq 1$ TeV while, as can be seen in the right panel of fig.~\ref{fig:isomchi_zg_tb50_nf1}.
The neutralino is a very pure bino
except in a very small region (which is in fact hardly visible in the figure) around the excluded zone in which
$Z_g<0.1$, i.e. a very pure higgsino.

\begin{figure}[t]
\centering
\includegraphics[scale=0.88]{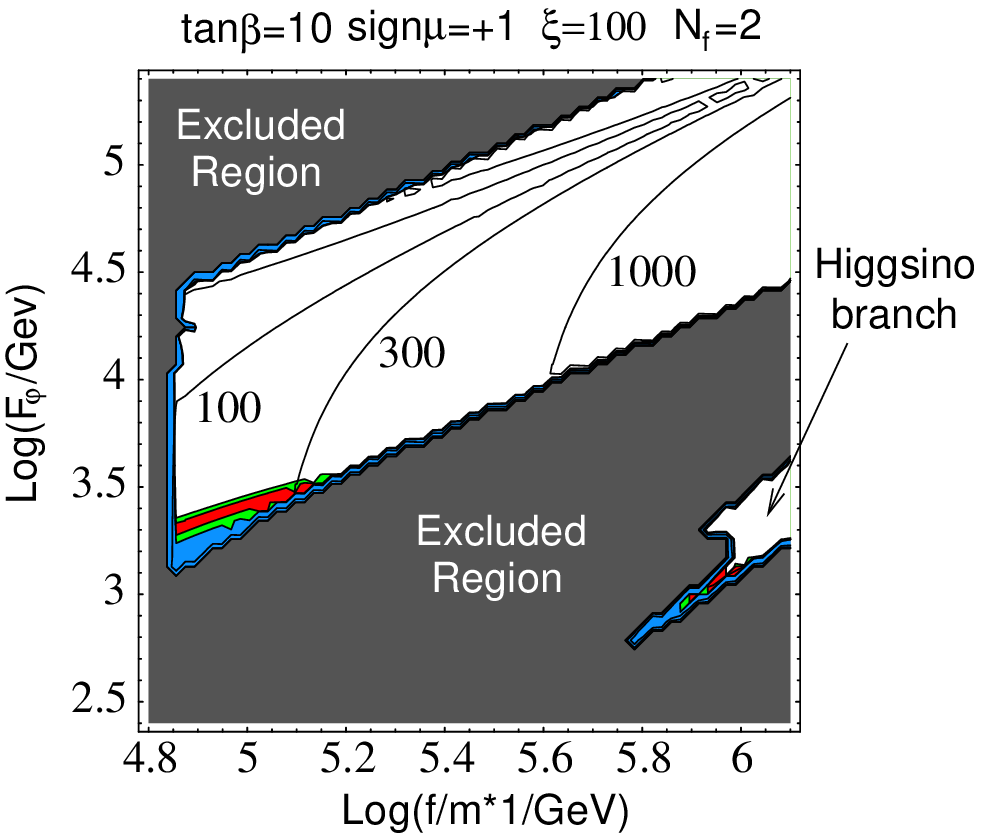}
\includegraphics[scale=0.88]{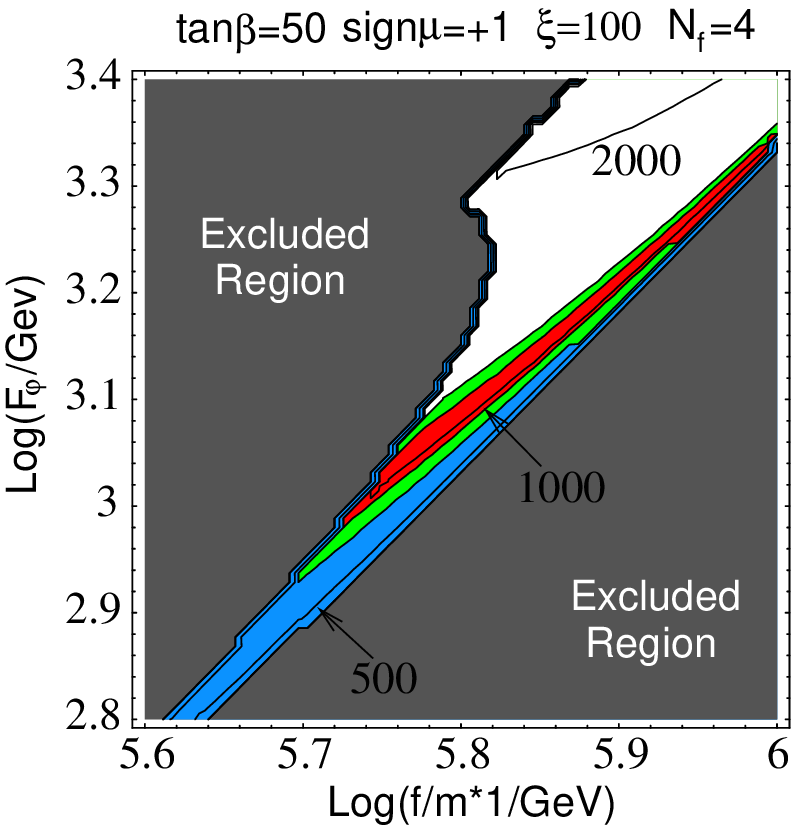}
\caption{Neutralino mass (expressed in GeV) contour plots in the plane
  $(f/m,F_\varphi)$ for $N_f>1$. Left panel: $N_f=2$ scenario. Right
  panel: higgsino region in the $N_f=4$ scenario.}
\label{fig:isomchi_nf2_nf4}
\end{figure}

The results for $N_f>1$ are shown in
fig.~\ref{fig:isomchi_nf2_nf4}. It is worth noting that in this case
there is a new branch in the parameter space in the region of high
$f/m$ (whose shape in general depends from $\xi$ and $\tan\beta$) in which the neutralino
is a very pure higgsino. In the right panel of
fig.~\ref{fig:isomchi_nf2_nf4} we show the higgsino region for
$\tan\beta=50$ and $N_f=4$. The cosmologically allowed region is
centered around $m_\chi\simeq1$ TeV, that is a quite natural result
for a heavy higgsino-like neutralino~\cite{Edsjo:1997bg,Chattopadhyay:2005mv}.
This is due the fact that in this case the $W^+ W^-$ channel (through
a chargino exchange) and the $Z^0 Z^0$ channel
are no longer suppressed and thus this implies a higher
annihilation cross section.

\section{Conclusions}
In this paper we showed that in the framework of the deflected
anomaly mediated scenario, which solves the problem of the
tachyons of the minimal anomaly mediation, the neutralino is still
the LSP in a wide region of the parameter space. This is achieved
considering the effects of the presence of a gauge mediated
sector. While in the standard anomaly mediation the neutralino is
wino like, this is no longer true for the scenario discussed here.
In fact the neutralino turns out to be a very pure bino or a very
pure higgsino depending on the number of messengers $N_f$ in the
gauge mediated sector. We have also computed the thermal relic
density (considering the standard cosmological scenario) and we
found that there are regions compatible with the latest WMAP data
both for $N_f=1$ and $N_f>1$.

\appendix
\section{Extraction of the Soft Terms \label{eotst}}
In this appendix we derive the soft supersymmetry breaking terms
starting from the lagrangian of eq. (\ref{lll}).
Let us start by writing it in a more compact form:
\begin{eqnarray}
L&=&\left[Z_Q\left(\rho^2+\delta\rho^2,\sigma^2+\delta\sigma^2\right)Q^\dagger Q\right]_{\theta^2{\bar{\theta}}^2}+\nonumber\\
& &+\left\{\left[y_0 Q^3+g^{-2}\left(\rho+\delta\rho,\sigma+\delta\sigma\right){\cal{W}}{\cal{W}}\right]_{\theta^2}+h.c.\right\},
\label{lllrhosigma}
\end{eqnarray}
where we have defined the quantities:
\begin{equation}
\rho=\frac{\mu}{m}
\label{ro}
\end{equation}
\begin{equation}
\delta\rho=-\rho\theta^2\frac{f}{m}
\label{dro}
\end{equation}
\begin{equation}
\sigma=\frac{m}{\Lambda_{UV}}
\label{sigma}
\end{equation}
\begin{equation}
\delta\sigma=\sigma\theta^2\left(\frac{f}{m}-F_\varphi\right)
\label{dsigma}
\end{equation}
\begin{equation}
\delta\rho^2=\rho^2\left(-\theta^2 \frac{f}{m}-{\bar{\theta}}^2\frac{f}{m}+\theta^2{\bar{\theta}}^2\frac{f^2}{m^2}\right)
\label{dro2}
\end{equation}
\begin{equation}
\delta\sigma^2=\sigma^2\left[\theta^2 \left(\frac{f}{m}-F_\varphi\right)+{\bar{\theta}}^2\left(\frac{f}{m}-F_\varphi\right)+\theta^2{\bar{\theta}}^2\left(\frac{f}{m}-F_\varphi\right)^2\right].
\label{dsigma2}
\end{equation}
After expanding eq. (\ref{lllrhosigma}) around\footnote{Note that the presence of the $\theta^2$, ${\bar{\theta}}^2$ in $\delta\rho$, $\delta\sigma$, $\delta\rho^2$, $\delta\sigma^2$ implies that the expansion ends at second order.} $(\rho,\sigma)$ and $(\rho^2, \sigma^2)$ , we can write the lagrangian as
\begin{eqnarray}
L&=&\left\{\left[Z_Q^i(\rho^2,\sigma^2)+Z_1^i(\rho^2,\sigma^2)\left(\theta^2+{\bar{\theta}}^2\right)+Z_2^i(\rho^2,\sigma^2)\theta^2{\bar{\theta}}^2\right]Q^\dagger_i Q_i\right\}_{\theta^2{\bar{\theta}}^2}+\nonumber\\
& &+\left\{\left[\frac{1}{6}y^{ijk} Q_i Q_j
    Q_k+\left(g^{-2}(\rho,\sigma)+\delta
      g^{-2}(\rho,\sigma)\theta^2\right){\cal{W}}{\cal{W}}\right]_{\theta^2}+\right.\nonumber\\
& &\left. +h.c.\right\},
\label{lllexp}
\end{eqnarray}
where we have posed:
\begin{eqnarray}
Z_1^i(\rho^2,\sigma^2) &=&-\frac{\partial Z_Q^i}{\partial \ln \rho^2} \frac{f}{m}+\frac{\partial Z_Q^i}{\partial \ln \sigma^2}\left(\frac{f}{m}-F_\varphi\right)\nonumber\\
Z_2^i(\rho^2,\sigma^2) &=&\frac{\partial Z_Q^i}{\partial \ln \rho^2} \frac{f^2}{m^2}+\frac{\partial Z_Q^i}{\partial \ln \sigma^2}\left(\frac{f}{m}-F_\varphi\right)^2+\nonumber\\
& &+\frac{\partial^2 Z_Q^i}{\partial^2 \ln
  \rho^2}\frac{f^2}{m^2}+\frac{\partial^2 Z_Q^i}{\partial^2 \ln
  \sigma^2}\left(\frac{f}{m}-F_\varphi\right)^2+\nonumber\\
& &-2\frac{f}{m}\left(\frac{f}{m}-F_\varphi\right)\frac{\partial^2 Z_Q^i}{\partial \ln \rho^2 \partial \ln \sigma^2}
\label{z123}
\end{eqnarray}
and
\begin{equation}
\delta g^{-2}(\rho,\sigma)=-\frac{\partial g^{-2}}{\partial \ln\rho}\frac{f}{m}+\frac{\partial g^{-2}}{\partial \ln\sigma}\left(\frac{f}{m}-F_\varphi\right).
\label{deltag-2}
\end{equation}
In eq.~(\ref{lllexp}) we have introduced the flavor indices in the kinetic and trilinear superpotential terms for later convenience.
In order to extract the soft supersymmetry breaking terms, we need to rescale the $Q_i$ and ${\cal{W}}$ chiral superfields in such a way that the new kinetic terms are canonically normalized.
The suitable definitions are:
\begin{equation}
g^{-1}{\cal{W}}={\cal{W}}^\prime
\label{wrescaling}
\end{equation}
\begin{equation}
{Z_Q^i}^{\frac{1}{2}}\left( 1+\frac{Z_1^i}{Z_Q^i}\theta^2\right)Q_i=Q_i^\prime,
\label{qrescaling}
\end{equation}
where $g^{-1}$ is evaluated at $(\rho,\sigma)$ and $Z_Q^i,Z_1^i$ at $(\rho^2,\sigma^2)$.
By expressing eq.~(\ref{lllexp}) in terms of the new (primed) fields, we obtain that:
\begin{eqnarray}
L &= &\left\{Q_i^{\prime\dagger}Q_i^\prime + \theta^2{\bar{\theta}}^2\left[\frac{Z_2^i}{Z_Q^i}-\frac{Z_1^{i2}}{Z_Q^{i2}}\right] Q_i^{\prime\dagger}Q_i^\prime\right\}_{\theta^2{\bar{\theta}}^2}+\nonumber\\
& &+\left\{\left[{\cal{W}}^\prime{\cal{W}}^\prime+\theta^2 \left(g^{2} \delta g^{-2}\right){\cal{W}}^\prime{\cal{W}}^\prime\right]_{\theta^2}\right\}+\nonumber\\
&
&+\left\{\left.\frac{1}{6}\frac{y^{ijk}}{\left({Z_Q^i}{Z_Q^j}{Z_Q^k}\right)^{\frac{1}{2}}}
    \left[1-\left(\frac{Z_1^{i}}{Z_Q^{i}}+\frac{Z_1^{j}}{Z_Q^{j}}+\frac{Z_1^{k}}{Z_Q^{k}}\right)\right]Q_i^\prime Q_j^\prime Q_k^\prime\right|_{\theta^2}+\right.\nonumber\\
& &\left. +h.c.\right\}.
\label{lllriscalata}
\end{eqnarray}
In our normalization (which is the same of~\cite{bw}) the lowest component of the gauge field strength is ${\cal{W}}^\prime=-i\frac{\lambda^\prime}{2}$.
By dropping the prime on all the fields and by redefining
\[
\frac{y^{ijk}}{\left({Z_Q^i}{Z_Q^j}{Z_Q^k}\right)^{\frac{1}{2}}}\rightarrow y^{ijk},
\]
we finally see that the lagrangian density with canonically normalized kinetic terms contains the following soft breaking interactions:
\begin{eqnarray}
L &\supset &-\left[\frac{Z_1^{i2}}{Z_Q^{i2}}-\frac{Z_2^i}{Z_Q^i}\right]{\tilde{q}}_i^* {\tilde{q}}_i-\frac{1}{2}\left[\frac{1}{2}g^2\delta g^{-2}\right]\lambda\lambda+\nonumber\\
& &-\frac{1}{6}\left[\left(\frac{Z_1^{i}}{Z_Q^{i}}+\frac{Z_1^{j}}{Z_Q^{j}}+\frac{Z_1^{k}}{Z_Q^{k}}\right)y^{ijk}\right]{\tilde{q}}_i{\tilde{q}}_j{\tilde{q}}_k+h.c.
\end{eqnarray}
By comparing with the lagrangian~(\ref{lsoft}), we can read off the soft supersymmetry breaking parameters:
\begin{eqnarray}
M_\lambda & = &\frac{1}{2}g^2\delta g^{-2}\nonumber\\
m_{ij}^2 & = & \left(\frac{Z_1^{i2}}{Z_Q^{i2}}-\frac{Z_2^i}{Z_Q^i}\right)\delta_{ij}\\
a_{ijk} & = &\left(\frac{Z_1^{i}}{Z_Q^{i}}+\frac{Z_1^{j}}{Z_Q^{j}}+\frac{Z_1^{k}}{Z_Q^{k}}\right)y^{ijk}\nonumber\\
b_{ij} & = &0\nonumber
\label{soffici}
\end{eqnarray}
Note that we are not providing any solution for the $\mu$ problem.
Now we derive explicit expressions for the soft breaking parameters in terms of low energy observable quantities.

\section{Explicit Expressions for the Soft Terms \label{eeftst}}
We need to compute the wave function renormalizations and the gauge couplings at an arbitrary renormalization scale $\mu$ of the order of the typical mass of the MSSM particles.
The one loop order RG equation for the running gauge coupling is given by
\begin{equation}
\frac{d}{d\ln \mu}g=-\frac{b}{16\pi^2}g^3,
\label{rgeqg}
\end{equation}
where $b$ is the appropriate one loop $\beta$-function coefficient at the scale $\mu$.
We need to integrate eq.~(\ref{rgeqg}) between the low energy scale $\mu$ and the UV cutoff scale $\Lambda_{UV}$.
We have to consider that if we denote with $b$ the $\beta$-function
coefficient at the low energy scale, then above the messenger mass
scale $m$ the $\beta$-function coefficient becomes $b^\prime=b-N_f$,
where $N_f$ is the number of flavors of messengers running into the
loops for $\mu>m$.
In passing through the threshold $m$ when integrating eq.~(\ref{rgeqg}), we match the values of the running gauge couplings above and below $m$.
The computation yields
\begin{equation}
g^{-2}\left(\rho,\sigma\right)=g^{-2}(\Lambda_{UV})+\frac{b-N_f}{8\pi^2}\ln \sigma +\frac{b}{8\pi^2}\ln \rho,
\label{ginted}
\end{equation}
where the variables $\rho$ and $\sigma$ are defined in eqs. (\ref{ro}) and (\ref{sigma}).
With a similar calculation and using the last result, we can integrate the RG equation of the wave function renormalization $Z_Q^i$.
This, in the limit of small Yukawa couplings with respect to $g^2$, is given by
\begin{equation}
\frac{d}{d \ln \mu}\ln Z_Q^i=\sum_{G_{Q_i}} \frac{c}{4\pi^{2}}g^2.
\label{rgeqz}
\end{equation}
In the last formula the sum is extended to all the gauge groups under which $Q^i$ is charged and $c$ is the relative quadratic Casimir.
The result of the integration can be written as
\[
\ln Z_Q^i\left(\rho^2,\sigma^2\right)=\ln Z_Q^i\left(\Lambda_{UV}\right)+
\]
\[
+\sum_{G_{Q_i}}\left\{\frac{2c}{b-N_f}\ln\left[\frac{g^{-2}(\Lambda_{UV})+\frac{b-N_f}{8\pi^2}\ln \sigma}{g^{-2}(\Lambda_{UV})}\right]+\right.
\]
\begin{equation}
\left.+\frac{2c}{b}\ln\left[\frac{g^{-2}(\Lambda_{UV})+\frac{b-N_f}{8\pi^2}\ln \sigma+\frac{b}{8\pi^2}\ln \rho}{g^{-2}(\Lambda_{UV})+\frac{b-N_f}{8\pi^2}\ln \sigma} \right]\right\}.
\label{zinted}
\end{equation}
To compute the soft breaking parameters listed in eq. (\ref{soffici}),
we need to use eqs.~(\ref{ginted}), (\ref{zinted}) into eqs.~(\ref{z123}), (\ref{deltag-2}) and plug the result into eq.~(\ref{soffici}).
The soft breaking parameters that arise from such computation are
\begin{eqnarray}
m^2_{ij}(\mu)& =& \delta_{ij}\sum_{G_i}\frac{2cg^4(\mu)}{(4\pi)^4}\left\{\left(\frac{f}{m}\right)^2 N_f\left[\zeta^2+\frac{N_f}{b}(1-\zeta^2)\right]+\right.\nonumber\\
& &+F_\varphi^2\left[b+N_f\left(\zeta^2-2+\frac{N_f}{b}(1-\zeta^2)\right)\right]+\nonumber\\
& &+\left.\frac{2fF_\varphi}{m}N_f(1-\zeta^2)\left(1-\frac{N_f}{b}\right)\right\}\nonumber\\
M_\lambda(\mu)& = &-\frac{g^2(\mu)}{(4\pi)^2}\left[N_f\frac{f}{m}+(b-N_f)F_\varphi\right]\\
a_{ijk}(\mu)&=&\sum_{G_i,G_j,G_k}\frac{2cg^2(\mu)}{(4\pi)^2}\left\{\frac{f}{m}\frac{N_f}{b}(\zeta-1)\right.
+\nonumber\\
& &-\left. F_\varphi\left[1+\frac{N_f}{b}(\zeta-1)\right]\right\}y_{ijk}\nonumber
\label{soffici2}
\end{eqnarray}
where we have defined
\begin{equation}
\zeta(\mu)=\left[1+\frac{b g^2(\mu)}{8\pi^2}\ln\left(\frac{m}{\mu}\right)\right]^{-1}=\left[1-\frac{b g^2(\mu)}{8\pi^2}\alpha(\mu)\ln\rho \right]^{-1}.
\label{xi}
\end{equation}
It is important to check that from eq. (\ref{soffici2}), in the limit in which the GMSB-like contribution to the soft terms decouples, we can recover the standard AMSB results for the soft terms.
This is easy to verify, since in the decoupling case\footnote{See the discussion in section~\ref{ssbt}.} $f/m=F_\varphi$ we get:
\bea
m^2_{ij}(\mu) & =& \delta_{ij}\sum_{G_i}\frac{2cg^4(\mu)}{(4\pi)^4}bF_\varphi^2 \nonumber\\
M_\lambda(\mu)& = &-\frac{g^2(\mu)}{(4\pi)^2}b F_\varphi\\
a_{ijk}(\mu) & = &-\sum_{G_i,G_j,G_k}\frac{2cg^2(\mu)}{(4\pi)^2}F_\varphi y_{ijk}\nonumber
\eea
which are the standard AMSB results of \cite{Randall:1998uk}.
In the phenomenological study we are interested in, we give the boundary conditions for the soft terms at the renormalization scale $\mu=m$, which corresponds to the typical messenger mass.
At such scale the results of eq.~(\ref{soffici2}) simplify, since $\zeta(m)=1$.
We can then write:
\bea
\left.m_{\lambda}\right|_{\mu=m}&=& -\frac{g^2(m)}{(4\pi)^2}\left[N_f\frac{f}{m}+(b-N_f)F_\varphi\right]\nonumber\\
\left.m^2_{ij}\right|_{\mu=m}&=&\delta_{ij}\sum_{G_i}\frac{2c g^4(m)}{(4\pi)^4}\left[\left(\frac{f}{m}\right)^2 N_f+F_\varphi^2\left(b-N_f\right)\right]\\
\left.a_{ijk}\right|_{\mu=m}&=&-\sum_{G_i,G_j,G_k}\frac{2c g^2(m)}{(4\pi)^2}F_\varphi y_{ijk}\nonumber
\label{soffici3}
\eea
We observe that in~(\ref{soffici3}) the soft terms at the messenger scale are exactly the sum of the AMSB-like and GMSB-like contributions.
In table {\ref{cb}} we report the values of the one loop $\beta$-function coefficients and the quadratic Casimirs for the standard model gauge groups and particles.
\begin{table}
\[
\begin{array}{|c|c|c|c|}
\hline
\mbox{Field} & c_1 & c_2 & c_3 \\
\hline
Q & 1/60 & 3/4 & 4/3 \\
t & 4/15 & \ & 4/3 \\
b & 1/15 & \ & 4/3 \\
L & 3/20 & 3/4 & \ \\
\tau & 3/5 & \ & \ \\
H_u & 3/20 & 3/4 & \ \\
H_d & 3/20 & 3/4 & \ \\
\hline
\end{array}
\hspace{.5cm}
\begin{array}{|c|c|}
\hline
\mbox{Group} & b \\
\hline
U(1)_Y & -33/5 \\
SU(2)_L & -1 \\
SU(3)_C &  3 \\
\hline
\end{array}
\]
\label{cb}
\caption{Quadratic Casimirs for the MSSM particles and one loop $\beta$-function coefficients for the gauge groups (the indices $1,2,3$ indicate the $U(1)_Y$, $SU(2)_L$, $SU(3)_C$ gauge groups respectively). }
\end{table}
To obtain our final predictions for the soft supersymmetry breaking
terms at the renormalization scale $m$ and in the limit of small
Yukawa couplings, we only have to put values contained in
table~\ref{cb} into eq.~(\ref{soffici3}).

The results are listed in the following equations.
\bea
\left.M_{1}\right|_{\mu=m}&=&-\frac{g_1^2(m)}{(4\pi)^2}\left[N_f\frac{f}{m}+(-\frac{33}{5}-N_f)F_\varphi\right]\nonumber\\
\left.M_{2}\right|_{\mu=m}&=&-\frac{g_2^2(m)}{(4\pi)^2}\left[N_f\frac{f}{m}+(-1-N_f)F_\varphi\right]\\
\left.M_{3}\right|_{\mu=m}&=&-\frac{g_3^2(m)}{(4\pi)^2}\left[N_f\frac{f}{m}+(3-N_f)F_\varphi\right]\nonumber
\label{mgauginiexpl}
\eea
\bea
\left.A_t\right|_{\mu=m}&=&-\frac{1}{(4\pi)^2}\left(\frac{16}{3}g_3^2+3g_2^2+\frac{13}{15}g_1^2\right)y_t\nonumber\\
\left.A_b\right|_{\mu=m}&=&-\frac{1}{(4\pi)^2}\left(\frac{16}{3}g_3^2+3g_2^2+\frac{7}{15}g_1^2\right)y_b\\
\left.A_\tau\right|_{\mu=m}&=&-\frac{1}{(4\pi)^2}\left(3g_2^2+\frac{9}{5}g_1^2\right)y_\tau\nonumber
\label{trilinearexpl}
\eea
\bea
\left.{\tilde{m}}_{Q}^2\right|_{\mu=m}&=&diag(1,1,1)\times\frac{2}{(4\pi)^4}\left\{\left[\left(\frac{f}{m}\right)^2 N_f+(3-N_f)F_\varphi^2\right]\frac{4}{3}g_3^4(m)\right.+\nonumber\\
& &+\left[\left(\frac{f}{m}\right)^2 N_f+(-1-N_f)F_\varphi^2\right]\frac{3}{4}g_2^4(m)+ \nonumber\\
& &\left.+\left[\left(\frac{f}{m}\right)^2 N_f+\left(-\frac{33}{5}-N_f\right)F_\varphi^2\right]\frac{1}{60}g_1^4(m)\right\}\nonumber\\
\left.{\tilde{m}}_{t}^2\right|_{\mu=m}&=&diag(1,1,1)\times\frac{2}{(4\pi)^4}\left\{\left[\left(\frac{f}{m}\right)^2 N_f+(3-N_f)F_\varphi^2\right]\frac{4}{3}g_3^4(m)\right.+\nonumber\\
& &\left.+\left[\left(\frac{f}{m}\right)^2 N_f+\left(-\frac{33}{5}-N_f\right)F_\varphi^2\right]\frac{4}{15}g_1^4(m)\right\}\nonumber\\
\left.{\tilde{m}}_{b}^2\right|_{\mu=m}&=&diag(1,1,1)\times\frac{2}{(4\pi)^4}\left\{\left[\left(\frac{f}{m}\right)^2 N_f+(3-N_f)F_\varphi^2\right]\frac{4}{3}g_3^4(m)\right.+\\
& &\left.+\left[\left(\frac{f}{m}\right)^2 N_f+\left(-\frac{33}{5}-N_f\right)F_\varphi^2\right]\frac{1}{15}g_1^4(m)\right\}\nonumber\\
\left.{\tilde{m}}_{L}^2\right|_{\mu=m}&=&diag(1,1,1)\times\frac{2}{(4\pi)^4}\left\{\left[\left(\frac{f}{m}\right)^2 N_f+(-1-N_f)F_\varphi^2\right]\frac{3}{4}g_2^4(m)\right.+ \nonumber\\
& &\left.+\left[\left(\frac{f}{m}\right)^2 N_f+\left(-\frac{33}{5}-N_f\right)F_\varphi^2\right]\frac{3}{20}g_1^4(m)\right\}\nonumber\\
\left.{\tilde{m}}_{\tau}^2\right|_{\mu=m}&=&diag(1,1,1)\times\frac{2}{(4\pi)^4}\left\{\left[\left(\frac{f}{m}\right)^2 N_f+\left(-\frac{33}{5}-N_f\right)F_\varphi^2\right]\frac{3}{5}g_1^4(m)\right\}\nonumber\\
\left.{m}_{H_u}^2\right|_{\mu=m}&=&\left.{m}_{H_d}^2\right|_{\mu=m}=\frac{2}{(4\pi)^4}\left\{\left[\left(\frac{f}{m}\right)^2 N_f+(-1-N_f)F_\varphi^2\right]\frac{3}{4}g_2^4(m)\right.+\nonumber\\
& &\left.+\left[\left(\frac{f}{m}\right)^2 N_f+\left(-\frac{33}{5}-N_f\right)F_\varphi^2\right]\frac{3}{20}g_1^4(m)\right\}\nonumber
\label{mscalexpl}
\eea

\end{document}